\documentstyle[]{article}
\font\cero=cmss10 scaled 1728 \font\uno=cmssbx10 scaled 1200
\begin{document}
\small{
\begin{flushleft}
{\cero  Deformation dynamics and the Gauss-Bonnet topological term in string theory} \\[3em]
\end{flushleft}
{\sf Alberto Escalante}\\
{\it Instituto de F\'{\i}sica, Universidad Aut\'onoma de Puebla,
Apartado postal J-48 72570, Puebla Pue., M\'exico.}
(aescalan@sirio.ifuap.buap.mx) \\[4em]

\noindent{\uno Abstract} \vspace{.5cm}\\
We show that there exist a nontrivial contribution on the Witten
covariant phase space when the Gauss-Bonnet topological term is
added to the Dirac-Nambu Goto action describing strings, because
of the geometry of deformations is modified, and on such space we
construct a symplectic structure. Future extensions of the present
results are outlined.

\noindent \\

\begin{center}
{\uno I. INTRODUCTION}
\end{center}
\vspace{1em} \ As we know, if we add the Gauss-Bonnet [GB]
topological term in any action describing strings (for example the
Dirac-Nambu-Goto action [DNG]), we do not find any contribution to
the equations of motion, because of the field equations of the
[GB] topological term are proportional to the called Einstein
tensor, and it does not give any contribution to the dynamics in a
two-dimension worldsheet swept out by a string, since the Einstein
tensor vanishes for such a geometry. In this manner, if we use the
conventional canonical formalism  based in the classical dynamics
of the system, we would not find apparently nothing
interesting.\\
However, using a covariant description of the canonical formalism
\cite{1}, and  identifying the arguments of the total divergences
at the level of the lagrangian as symplectic potentials \cite{2},
in \cite{3} Cartas-Fuentevilla gives a sign that the [GB]
topological term has a nontrivial contribution in the symplectic
structure constructed on the classical covariant phase space. But,
in \cite{3} important calculations are not developed, for example:
the contribution of the [GB] topological term to the linearized
equations of motion that are useful for stability analisis,
subsequently the construction of a covariantly conserved
symplectic current that allows us to stablish a conection between
functions and  Hamiltonian vector fields, and
 the correct
identification of the exterior derivative
on the phase space.\\
In this manner, the purpose of this article is to show in a clear
way how the [GB] topological term modifies the symplectic geometry
of the Witten covariant phase space, when we add it to the [DNG]
action for string theory, by developing new ideas and completing
the results presented
in \cite{3}. \\
This paper is organized as follows. In the Sect. II, using the
deformations formalism introduced in \cite{4}, we calculate the
normal and tangential deformations of quantities on the embedding,
that will be useful in our developments. In Sect. III, we obtain
the equations of motion and identify the corresponding symplectic
potential for [DNG-GB] p-branes, also we calculate the linearized
equations of motion and show that in general the [GB] topological
term gives indeed a nontrivial contribution on the deformations
geometry. In Sect. III.I, we take the results obtained in Sect.III
for the case of string theory, and we show that in spite of the
dynamics for [DNG] and [DNG-GB] in string theory are the same, the
simplectic potential, and the linearized dynamics are modified
because of the [GB] topological term, and therefore, there is a
relevant contribution on the phase space. In Sect. IV, from the
linearized equations obtained in Sect. III.I for string theory, we
obtain a symplectic current by applying the self-adjoint operators
scheme, proving that is a world sheet covariantly conserved
current. In Sect. V, we define the Witten covariant phase space
for [DNG-GB] strings, and using the symplectic current found in
Sect. IV, we construct a geometrical structure, showing that it
has a relevant contribution because of the [GB] topological term.
In Sect. VI, we give conclusions and prospects.

\setcounter{equation}{0} \label{c2}.

\noindent \textbf{II. Deformations of the embedding}\\[1ex]
In the scheme of deformations \cite{4}, the physically observable
measure of the deformations of the embedding is given by the
orthogonal projection of the infinitesimal spacetime variations
$\delta X^{\mu}=n{_{i}}^{\mu}  \phi^{i}$, and the tangential
deformations together with the total divergence terms are
neglected. However, In \cite{2} it is  shown that tangential
deformations and divergence terms are important because of  such
terms are identified as symplectic potentials, whose  variations
(the exterior derivative on the space phase) generate the integral
kernel of a covariant and gauge invariant symplectic structure,
for the theory under study. Thus, for a complete analysis, we do
not only  need calculate the normal deformations of fields defined
on the embedding as in \cite{4},
but also  the tangential deformations, that will be important in the developed of this work. \\
For this purpose, we decompose an arbitrary infinitesimal
deformation of the embedding $\delta X^{\mu}$ into its parts
tangential and normal to the worldsheet
\begin{equation}
\delta X^{\mu}=e{_{a}}^{\mu} \phi^{a} + n{_{i}}^{\mu} \phi^{i},
\end{equation}
where $ n{_{i}}^{\mu}$ are vector fields normal to the worldsheet and $e{_{a}}^{\mu}$
are vector fields tangent to such a worldsheet, thus,
the deformation operator is defined as
\begin{equation}
\textbf{D}=D_{\delta} + D_{\Delta},
\end{equation}
 where
\begin{equation}
D_{\delta}= \delta^{\mu}D_{\mu}, \quad \quad
\delta^{\mu}= n{_{i}}^{\mu} \phi^{i},
\end{equation}
and
\begin{equation}
D_{\Delta}= \Delta^{\mu} D_{\mu}, \quad \quad \Delta^{\mu}= e{_{a}}^{\mu}
\phi^{a},
\end{equation}
therefore, we can find that the deformations of the intrinsic
geometry of the embedding are given by
\begin{equation}
\textbf{D}e_{a}= (K{_{ab}}^{i} \phi_{i})e^{b} +
(\widetilde \nabla_{a}\phi_{i}) n^{i} + (\nabla_{a}
\phi^{b})e_{b} - K{_{ab}}^{i} \phi^{b} n_{i},
\end{equation}
\begin{equation}
\textbf{D} \gamma_{ab}= 2K{_{ab}}^{j} \phi_{j} +
\nabla_{a} \phi_{b}+\nabla_{b} \phi_{a},
\end{equation}
\begin{equation}
\textbf{D}\gamma^{ab}= -2K^{abj} \phi_{j} -
\nabla^{a} \phi^{b}-\nabla^{b} \phi^{a},
\end{equation}

\begin{equation}
\textbf{D}\sqrt{- \gamma}= \sqrt{-
\gamma}[\nabla_{a}\phi^{a}+ K^{i} \phi_{i}],
\end{equation}
\begin{eqnarray}
\nonumber \textbf{D} \gamma{_{gf}}^{a} \!\!\! & = &
\!\!\!
\gamma^{ad}[ \nabla_{f} (K{_{gd}}^{j} \phi_{j}) + \nabla_{g} (K{_{fd}}^{j} \phi_{j})- \nabla_{d}((K{_{gf}}^{j} \phi_{j})] \nonumber \\
\!\!\!  & + & \!\!\!  \frac{1}{2} \gamma^{ad}[2
\nabla_{(g}\nabla_{f)} \phi_{d} -R^{e}{_{fdg}} \phi_{e}-
R^{e}{_{gdf}} \phi_{e}],
\end{eqnarray}
\begin{equation}
\textbf{D} R_{ab}=\nabla_{c}(\textbf{D}
\gamma{_{ab}}^{c})- \nabla_{b}(\textbf{D}
\gamma{_{ac}}^{c}),
\end{equation}
\begin{eqnarray}
\nonumber \textbf{D} R \!\!\! & = & \!\!\!  (\textbf{D} \gamma^{ab})R_{ab} +  \gamma^{ab}(\textbf{D} R_{ab}) \nonumber \\
\!\!\! & = & \!\!\! -2 \nabla^{a}\phi^{b}R_{ab}- 2
K^{abj}\phi_{j} R_{ab} +
\gamma^{ab}[\nabla_{c}(\textbf{D}\gamma{_{ab}}^{c})-
\nabla_{b}(\textbf{D} \gamma{_{ac}}^{c}) ],
\end{eqnarray}
where, $K{_{ab}}^{i}$, $ \gamma_{ab}$, $\gamma{_{gf}}^{a}$, $R$,
$R_{ab}$ are the extrinsic curvature, the metric, the connection
coefficients, the scalar curvature and the Ricci tensor of the
world-sheet respectively \cite{4}.\\
For this article this is sufficient on the deformations of embedding.\\
\\
\\

\noindent \textbf{III. The DNG action for p-branes  with a Gauss-Bonnet term }\\[1ex]
As we know, the [DNG] action for p-branes is proportional to the
area of the spacetime trayectory created by the brane, and the
Gauss-Bonnet term is proportional to the Ricci scalar
$R$ constructed from the world surface metric $\gamma_{ab}$. Both
terms are given in the following action
\begin{equation}
S= -  \sigma \int  \sqrt{ -\gamma}d^{D}\xi + \beta \int  \sqrt{
-\gamma} R d^{D}\xi,
\end{equation}
where $\sigma$ y $\beta$ are constants, and $D$ is the dimension
of the world sheet.
\\
In agreement  with (8)-(11), the variation of the action (12) is
given by
\begin{equation}
\textbf{D} S = - \int  \sqrt{ -\gamma}[ \sigma K^{i}
+ 2 \beta G_{ab}K^{abi}] \phi_{i}  d^{D}\xi+ \int
\sqrt{ -\gamma} \nabla_{a}[- \sigma \phi^{a}-2 \beta
G^{ab}\phi_{b} + \beta \gamma^{cd} \textbf{D}
\gamma_{{cd}}^{a}- \beta \gamma ^{ab}\textbf{D}
\gamma{_{cb}}^{c}]d^{D}\xi,
\end{equation}
where we can identify the equations of motion
\begin{equation}
\sigma K^{i} +  2\beta G_{ab}K^{abi}=0,
\end{equation}.
being $G_{ab}$ the world surface Einstein tensor given by
\begin{equation}
G_{ab}= R_{ab}- \frac{1}{2} \gamma_{ab} R,
\end{equation}
and the  argument of the pure divergence term is identified as a
symplectic potential for the theory \cite{2}, as it will be proved
below.
\begin{equation}
\Psi^{a}= \sqrt{ -\gamma} [- \sigma \phi^{a}- 2\beta
G^{ab}\phi_{b} + \beta \gamma^{cd} \textbf{D} \gamma_{{cd}}^{a}-
\beta \gamma ^{ab}\textbf{D} \gamma{_{cb}}^{c}].
\end{equation}
We notice from equation (16), that the symplectic potential found
in \cite{3} is incomplete; the reason is that the normal variation
$(D_{\delta})$ is considered as exterior derivative on the phase
space, however, as we will show in the next sections the correct
exterior derivative on the phase space is the sum of
normal and tangential variations $(D_{\delta}+D_{\Delta})$.\\
On the other hand, such as in \cite{2}, the variation of $\Psi^{a}$
(the derivative exterior on the phase space) will
generate the integral kernel of a covariant and gauge invariant
symplectic structure for [DNG-GB] theory. In this manner, we can
see in equation (16) that in general there is a relevant
contribution on the phase space because of the terms proportional to
the parameter $\beta$, coming from the [GB] term. \\
In order to give a more detailed analysis, let us see how the [GB]
term contribute to the linearized equations of motion when we add
it to the [DNG] action for p-branes, for this we calculate the
variations of the equation (14), obtaining
\begin{eqnarray}
 \nonumber \!\!\!\ & \sigma & \!\!\!\  [-\widetilde \Delta^{i}_{j}-
K{_{ab}}^{i}K{^{ab}}_{j} + g(\textbf{R}(e_{a},n_{j})e^{a},n^{i}) ] \phi^{j}
+ 2 \beta G^{ab}[ - \widetilde \nabla_a \widetilde \nabla_{b}
\phi^{i} + K{_{ad}}^{i}K{^{d}}_{b}{^{j}} \phi_{j} +
g(\textbf{R}(e_{a},n^{j})e_{b},n^{i}) \phi_{j}] \nonumber \\ \!\!\!\ & - &
\!\!\!\ 8 \beta K{^{b}}_{d}{^{i}}K^{adj} \phi_{j} G_{ab} + 4 \beta
K^{abi} \widetilde \nabla_{c} \widetilde \nabla_{b} K{_{a}}^{cj}
\phi_{j} + 4 \beta K^{abi} \widetilde \nabla_{b} K{_{a}}^{cj}
\widetilde \nabla_{c} \phi_{j} + 4 \beta K^{abi} \widetilde
\nabla_{c} K{_{a}}^{cj} \widetilde \nabla_{b} \phi_{j}
\nonumber \\
\!\!\! & + & \!\!\! 4 \beta K^{abi} K{_{a}}^{cj}\widetilde
\nabla_{c} \widetilde \nabla_{b} \phi_{j} - 2 \beta K^{abi}
\widetilde \Delta K{_{ab}}^{j} \phi_{j} -4 \beta K^{abi}
\widetilde \nabla_{c} K{^{ab}}^{j} \widetilde \nabla^{c} \phi_{j}
- 2 \beta K^{abi}K{_{ab}}^{j} \widetilde \Delta \phi_{j} - 2 \beta
K^{abi} \widetilde \nabla_{b} \widetilde \nabla_{a}K^{j} \phi_{j}
\nonumber \\ \!\!\! & - & \!\!\!  4 \beta K^{abi} \widetilde
\nabla_{a}K^{j} \widetilde \nabla_{b} \phi_{j} - 2 \beta
K^{abi}K^{j} \widetilde \nabla_{b} \widetilde \nabla_{a} \phi_{j}
-2 \beta  K^{abi} K{_{ab}}^{j} \phi^{j}R + 2 \beta
R_{cd}K{^{cd}}_{j} \phi^{j} K^{i} + 2 \beta K^{i} \widetilde
\Delta K^{j} \phi_{j} \nonumber \\ \!\!\!\ & + & \!\!\!\ 4 \beta
K^{i} \widetilde \nabla_{c} K^{j} \widetilde \nabla^{c} \phi_{j}+
2 \beta K^{i}K^{j} \widetilde \Delta \phi_{j} - 2 \beta K^{i}
\widetilde \nabla_{c} \widetilde \nabla_{g} K{^{gcj}} \phi_{j} - 4
\beta K^{i} \widetilde \nabla_{g}K^{gcj} \widetilde \nabla_{c}
\phi_{j} \nonumber \\
\!\!\!\ & - & \!\!\!\ 2 \beta K^{i}K^{cgj} \widetilde \nabla_{c}
\widetilde \nabla_{g}\phi_{j}=0,
\end{eqnarray}
where $ g(\textbf{R}(Y_{1},Y_{2},)Y_{3},Y_{4}) \equiv
\textbf{R}_{\alpha \beta \gamma
\nu}Y{_{2}}^{\alpha}Y{_{1}}^{\beta}Y{_{3}}^{\gamma}
Y{_{4}}^{\nu}$, being $\textbf{R}_{\alpha \beta \gamma \nu}$ the
background Riemman tensor \cite{4}. We can identify the first term
of last equation as the linearized dynamics of [DNG] theory, and
the proportional terms to the parameter $\beta$ as the
contribution of [GB] term. As one would expect, if the parameter
$\beta$ vanished, we should obtain the linearized equations for
[DNG] theory [4,5]. Thus, we can see that in general there is a
relevant contribution to linearized equations because
of [GB] term when we add it in the [DNG] action for p-branes. \\

\noindent \textbf{III.I  The DNG action with a Gauss-Bonnet topological term in closed string theory}\\[1ex]
In this section we will see what happens when we consider in
equations (14), (16), (17), the case of string theory. For this,
we know that in a two-dimensional world sheet surface, swept out
for a string, the Einstein tensor vanishes $G_{ab}=0$, and
equation (14) takes the form
\begin{equation}
K^{i}=0,
\end{equation}
where we can see that the dynamics for  [DNG] and  [DNG-GB] in
string theory are the same, and we do not find any contribution
because of [GB] topological term, thus, if we use a conventional
formulation to quantize the [DNG-GB] strings from corresponding
classical dynamics (equation (18)) the same result is obtained and
we would not find apparently any interest for including the [GB]
topological term in any action describing strings. However,
whether in equation (16) we consider the case of string theory we
obtain
\begin{equation}
\Psi^{a}= \sqrt{ -\gamma} [- \sigma \phi^{a} + \beta \gamma^{cd} \textbf{D} \gamma_{{cd}}^{a}-
\beta \gamma ^{ab}\textbf{D} \gamma{_{cb}}^{c}],
\end{equation}
in this manner, we can see that the last two terms
correspond to the topological term that do not vanish and
 give a nontrivial contribution to the phase space as we will see in the next sections. \\
For the purposes of this paper, we take the case of string theory
in equation (17), obtaining
\begin{eqnarray}
 \nonumber \!\!\!\ & \sigma & \!\!\!\  [-\widetilde \Delta^{i}_{j}-
K{_{ab}}^{i}K{^{ab}}_{j} + g(\textbf{R}(e_{a},n_{j})e^{a},n^{i}) ] \phi^{j}
 \nonumber \\ \!\!\!\ & + & \!\!\!\ 4 \beta K^{abi} \widetilde
\nabla_{c} \widetilde \nabla_{b} K{_{a}}^{cj} \phi_{j} + 4 \beta
K^{abi} \widetilde \nabla_{b} K{_{a}}^{cj} \widetilde \nabla_{c}
\phi_{j} + 4 \beta K^{abi} \widetilde \nabla_{c} K{_{a}}^{cj}
\widetilde \nabla_{b} \phi_{j}
\nonumber \\
\!\!\! & + & \!\!\! 4 \beta K^{abi} K{_{a}}^{cj}\widetilde
\nabla_{c} \widetilde \nabla_{b} \phi_{j} - 2 \beta K^{abi}
\widetilde \Delta K{_{ab}}^{j} \phi_{j} -4 \beta K^{abi}
\widetilde \nabla_{c} K{^{ab}}^{j} \widetilde \nabla^{c} \phi_{j}
- 2 \beta K^{abi}K{_{ab}}^{j} \widetilde \Delta \phi_{j} - 2 \beta
K^{abi} \widetilde \nabla_{b} \widetilde \nabla_{a}K^{j} \phi_{j}
\nonumber \\ \!\!\! & - & \!\!\!  4 \beta K^{abi} \widetilde
\nabla_{a}K^{j} \widetilde \nabla_{b} \phi_{j} - 2 \beta
K^{abi}K^{j} \widetilde \nabla_{b} \widetilde \nabla_{a} \phi_{j}
-2 \beta  K^{abi} K{_{ab}}^{j} \phi^{j}R + 2 \beta
R_{cd}K{^{cd}}_{j} \phi^{j} K^{i} + 2 \beta K^{i} \widetilde
\Delta K^{j} \phi_{j} \nonumber \\ \!\!\!\ & + & \!\!\!\ 4 \beta
K^{i} \widetilde \nabla_{c} K^{j} \widetilde \nabla^{c} \phi_{j}+
2 \beta K^{i}K^{j} \widetilde \Delta \phi_{j} - 2 \beta K^{i}
\widetilde \nabla_{c} \widetilde \nabla_{g} K{^{gcj}} \phi_{j} - 4
\beta K^{i} \widetilde \nabla_{g}K^{gcj} \widetilde \nabla_{c}
\phi_{j} \nonumber \\
\!\!\!\ & - & \!\!\!\ 2 \beta K^{i}K^{cgj} \widetilde \nabla_{c}
\widetilde \nabla_{g}\phi_{j}=0,
\end{eqnarray}
where we can see that there is also a contribution to the [DNG]'s
linearized equations because of [GB] topological term, which is
completely unknown in the literature. It is remarkable to mention
that the linearized equations for [DNG-GB] strings theory,equation
(20), can be useful in stability analysis, however, this is far
from our purposes and we shall leave it as an open question, and we
shall focus on the effects
of the [GB] topological term on the phase space.  \\
On the other hand, we consider the equation (18) in equation (20),
obtaining
\begin{equation}
P{^{ij}} \phi_{j}=0
\end{equation}
where the operator $P{^{ij}}$ is given for
\begin{eqnarray}
\nonumber \!\!\!\ P^{ij} & = & \!\!\!\  [ \sigma\{-\widetilde
\Delta^{ij}- K{_{ab}}^{i} K^{abj} + g(R(e_{a},n^{j})e^{a},n^{i})
\} + 4 \beta K^{abi} \widetilde \nabla_{c} \widetilde \nabla_{b}
K{_{a}}^{cj} + 4 \beta K^{abi} \widetilde \nabla_{b} K{_{a}}^{cj}
\widetilde \nabla_{c} \nonumber \\
\!\!\!\ & + & \!\!\!\ 4 \beta K^{abi} \widetilde \nabla_{c}
K{_{a}}^{cj} \widetilde \nabla_{b}
 + 4 \beta K^{abi} K{_{a}}^{cj} \widetilde \nabla_{c} \widetilde
\nabla_{b}  - 2 \beta K^{abi} \widetilde \Delta K{_{ab}}^{j}  -4
\beta K^{abi} \widetilde \nabla_{c}
K{^{ab}}^{j} \widetilde \nabla^{c}  \nonumber \\
\!\!\!\ & - & \!\!\! 2 \beta K^{abi}K{_{ab}}^{j} \widetilde
\Delta-2 \beta K^{abi} K{_{ab}}^{j}R].
\end{eqnarray}
In the next section we will apply the self-adjoint operators
method to equation (21), and  we will demonstrate that the
operator $P^{ij}$ given in equation (22) is self-adjoint,
obtaining a symplectic current from  this property.
\newline
\newline
\noindent \textbf{IV. Self-adjointness of the linearized dynamics}\\[1ex]
In this section we shall demonstrate that the operator
$P{^{i}}_{j}$, given in equation (22),  is indeed self-adjoint and
in this manner we shall construct a symplectic current in terms of
solutions of the equation (21). With this purpose, let
$\phi{_{1}}^{i}$ and $\phi{_{2}}^{i}$ be two arbitrary scalar
fields, which correspond to a pair of solutions of the equation
(21), thus we can verify the following:
\begin{equation}
 - \sigma \phi_{1i}\widetilde \Delta \phi{_{2}}^{i}= - \sigma
\widetilde \Delta \phi_{1i} \phi{_{2}}^{i} +
\nabla_{a}j{_{1}}^{a},
\end{equation}
\begin{eqnarray}
\nonumber \\ 4 \beta K^{abi} \widetilde \nabla_{c} \widetilde
\nabla_{b} K{_{a}}^{cj}\phi_{1i} \phi{_{2j}} \!\!\!\ & = & \!\!\!\
4 \beta K^{abi} \widetilde \nabla_{c} \widetilde \nabla_{b}
K{_{a}}^{cj}\phi_{1i} \phi{_{2j}}-4 \beta K^{abj} \widetilde
\nabla_{c} \widetilde \nabla_{b} K{_{a}}^{ci}\phi_{1i} \phi{_{2j}}
\nonumber \\ \!\!\!\ & + & \!\!\!\ 4 \beta K^{abj} \widetilde
\nabla_{c} \widetilde \nabla_{b} K{_{a}}^{ci}\phi_{1i}
\phi{_{2j}},
\end{eqnarray}
\begin{eqnarray}
\nonumber \\ 4 \beta K^{abi} \widetilde \nabla_{b} K{_{a}}^{cj}
\phi_{1i}\widetilde \nabla_{c} \phi_{2j} \!\!\!\ & = & \!\!\!\
-4\beta \widetilde \nabla_{a} K^{bci} \widetilde \nabla_{b}
K{_{c}}^{aj}\phi_{1i}\phi_{2j} - 4\beta K^{bci}\widetilde
\nabla_{a}\widetilde \nabla_{b}K{_{c}}^{aj}\phi_{1i}\phi_{2j}
\nonumber \\ \!\!\!\ & - & \!\!\!\ 4 \beta K^{bci}\widetilde
\nabla_{b}K{_{c}}^{aj} \widetilde \nabla_{a}\phi_{1i} \phi_{2j} +
\nabla_{a} j{_{2}}^{a},
\end{eqnarray}
\begin{eqnarray}
\nonumber \\ 4 \beta K^{abi} \widetilde \nabla_{c} K{_{b}}^{cj}
\phi_{1i} \widetilde \nabla_{a}\phi_{2j} \!\!\!\ & = & \!\!\!\ - 4
\beta \widetilde \nabla_{a}K^{abi}\widetilde
\nabla_{c}K{_{b}}^{cj}\phi_{1i}\phi_{2j}- 4 \beta K^{abi}
\widetilde \nabla_{a} \widetilde \nabla_{c} K{_{b}}^{cj} \phi_{1i}
\phi_{2j} \nonumber \\
\!\!\!\ & - & \!\!\!\ 4 \beta K^{abi} \widetilde \nabla_{c}
K{_{b}}^{cj} \widetilde \nabla_{a} \phi_{1i} \phi_{2j} +
\nabla_{a} j{_{3}}^{a},
\end{eqnarray}
\begin{eqnarray}
\nonumber \\ 4 \beta K^{abi} K{_{a}}^{cj} \phi_{1i}\widetilde
\nabla_{c} \widetilde \nabla_{b} \phi_{2j} \!\!\!\ & = & \!\!\!\
4\beta\widetilde \nabla_{a} \widetilde \nabla_{b} K^{cai}
K{_{c}}^{bj} \phi_{1i} \phi_{2j}+ 4 \beta \widetilde \nabla_{b}
K^{cai} \widetilde \nabla_{a} K{_{c}}^{bj} \phi_{1i} \phi_{2j}
\nonumber \\
\!\!\!\ & + & \!\!\!\ 4 \beta \widetilde \nabla_{b} K^{cai}
 K{_{c}}^{bj} \widetilde \nabla_{a} \phi_{1i} \phi_{2j} + 4 \beta \widetilde \nabla_{a} K^{cai}
 \widetilde \nabla_{b} K{_{c}}^{bj}  \phi_{1i} \phi_{2j} \nonumber \\  \!\!\!\ &
 + & \!\!\!\ 4 \beta  K^{cai}
 \widetilde \nabla_{a} \widetilde \nabla_{b} K{_{c}}^{bj}  \phi_{1i}
 \phi_{2j} + 4 \beta  K^{cai}
 \widetilde \nabla_{b} K{_{c}}^{bj}   \widetilde \nabla_{a} \phi_{1i}
 \phi_{2j} \nonumber \\
 \!\!\!\ & + & \!\!\!\ 4 \beta \widetilde \nabla_{a}  K^{cai}
  K{_{c}}^{bj} \widetilde \nabla_{b} \phi_{1i}
 \phi_{2j} + 4 \beta  K^{cai}
   \widetilde \nabla_{a} K{_{c}}^{bj} \widetilde \nabla_{b} \phi_{1i}
 \phi_{2j} \nonumber \\ \!\!\!\ & + & \!\!\!\  4 \beta  K^{cai}
    K{_{c}}^{bj} \widetilde \nabla_{a} \widetilde \nabla_{b} \phi_{1i}
 \phi_{2j} + \nabla_{a}j{_{4}}^{a},
\end{eqnarray}
\begin{eqnarray}
\nonumber \\ - 2 \beta K^{abi} \widetilde \Delta
K{_{ab}}^{j}\phi_{1i}
 \phi_{2j} \!\!\!\ & = & \!\!\!\  - 2 \beta K^{abi} \widetilde \Delta
K{_{ab}}^{j}\phi_{1i}
 \phi_{2j} + 2 \beta K^{abj} \widetilde \Delta
K{_{ab}}^{i}\phi_{1i}
 \phi_{2j} \nonumber \\
 \!\!\!\ & - & \!\!\!\ 2 \beta K^{abj} \widetilde \Delta
K{_{ab}}^{i}\phi_{1i}
 \phi_{2j},
\end{eqnarray}
\begin{eqnarray}
\nonumber \\ - 4 \beta K^{cdi} \widetilde \nabla_{a} K{_{cd}}^{j}
\phi_{1i}\widetilde \nabla^{a}\phi_{2j} \!\!\!\ & = & \!\!\!\ 4
 \beta \widetilde \nabla^{a} K^{cdi} \widetilde \nabla_{a}
K{_{cd}}^{j} \phi_{1i}\phi_{2j} + 4 \beta K^{cdi} \widetilde
\Delta K{_{cd}}^{j} \phi_{1i}\phi_{2j} \nonumber \\
\!\!\!\ & + & \!\!\!\ 4 \beta K^{cdi} \widetilde \nabla_{a}
K{_{cd}}^{j} \widetilde \nabla^{a} \phi_{1i}\phi_{2j} +
\nabla_{a}j{_{5}}^{a},
\end{eqnarray}
\begin{eqnarray}
\nonumber \\ - 2 \beta K^{cdi} K{_{cd}}^{j} \phi_{1i}\widetilde
\Delta \phi_{2j} \!\!\!\ & = & \!\!\!\ - 2 \beta  \widetilde
\Delta K^{cdi} K{_{cd}}^{j} \phi_{1i} \phi_{2j} -4 \beta
\widetilde \nabla_{a}K^{cdi} \widetilde
\nabla^{a}K{_{cd}}^{j}\phi_{1i} \phi_{2j} \nonumber \\  \!\!\!\ &
- & \!\!\!\ 4 \beta \widetilde \nabla_{a}K^{cdi} K{_{cd}}^{j}
\widetilde \nabla^{a} \phi_{1i} \phi_{2j}- 2 \beta K^{cdi}
\widetilde \Delta K{_{cd}}^{j} \phi_{1i} \phi_{2j} \nonumber \\
\!\!\!\ & - & \!\!\!\ 4 \beta K^{cdi} \widetilde \nabla_{a}
K{_{cd}}^{j} \widetilde \nabla^{a} \phi_{1i} \phi_{2j} - 2 \beta
K^{cdi} K{_{cd}}^{j} \widetilde \Delta \phi_{1i} \phi_{2j}
\nonumber \\ \!\!\!\ & + & \!\!\!\ \nabla_{a}j{_{6}}^{a},
\end{eqnarray}
 where
\begin{equation}
 j{_{1}}^{a}= \sigma [-\phi_{1i}\widetilde \nabla^{a}
\phi{_{2}}^{i} + \widetilde \nabla^{a} \phi_{1i} \phi{_{2}}^{i}],
\end{equation}
\begin{equation}
j{_{2}}^{a}=4 \beta K^{bci} \widetilde \nabla_{b}
K{_{c}}^{aj}\phi_{1i}\phi_{2j},
\end{equation}
\begin{equation}
j{_{3}}^{a}= 4 \beta K^{abi} \widetilde
\nabla_{c}K{_{b}}^{cj}\phi_{1i}\phi_{2j},
\end{equation}
\begin{eqnarray}
j{_{4}}^{a}= \beta [4 K^{cbi}K{_{c}}^{aj} \phi_{1i} \widetilde
\nabla_{b} \phi_{2j} - 4 \widetilde \nabla_{b} K^{cai}
K{_{c}}^{bj}\phi_{1i}\phi_{2j} - 4K^{cai} \widetilde \nabla_{b}
K{_{c}}^{bj}\phi_{1i}\phi_{2j} - 4 K^{cai}K{_{c}}^{bj} \widetilde
\nabla_{b}\phi_{1i}\phi_{2j}],
\end{eqnarray}
\begin{equation}
j{_{5}}^{a}= -4 \beta K^{cdi} \widetilde
\nabla^{a}K{_{cd}}^{j}\phi_{1i}\phi_{2j}
\end{equation}
\begin{equation}
j{_{6}}^{a}= \beta [-2 K^{cdi}K{_{cd}}^{j}\phi_{1i} \widetilde
\nabla^{a} \phi_{2j} + 2\widetilde \nabla^{a} K^{cdi}
K{_{cd}}^{j}\phi_{1i}\phi_{2j}+ 2 K^{cdi} \widetilde \nabla^{a}
K{_{cd}}^{j} \phi_{1i}\phi_{2j} + 2K^{cdi} K{_{cd}}^{j} \widetilde
\nabla^{a} \phi_{1i}\phi_{2j}].
\end{equation}
In this manner, considering the equations (22)-(35), and after
some arrangements, we obtain
\begin{equation}
\phi_{1i} (P^{ij}) \phi_{2j}=(P^{ji}) \phi_{1i} \phi_{2j}
+\nabla_{a}j^{a},
\end{equation}
where $j^{a}= \sum_{{i=1}}^{6} j{_{i}}^{a}$, which we can
simplifly by substituting explicitly   the equations (31)-(36):
\begin{eqnarray}
\nonumber \\ j^{a} \!\!\!\ & = & \!\!\!\  \sigma
[-\phi_{1i}\widetilde \nabla^{a} \phi{_{2}}^{i} + \widetilde
\nabla^{a} \phi_{1i} \phi{_{2}}^{i}] + 4 \beta K^{bci} \widetilde
\nabla_{b} K{_{c}}^{aj} \phi_{1i} \phi_{2j} -4 \beta K^{cdi}
\widetilde
\nabla^{a} K{_{cd}}^{j}\phi_{1i}\phi_{2j} \nonumber \\
\!\!\!\ & + & \!\!\!\ \beta [4 K^{cbi} K{_{c}}^{aj} \phi_{1i}
\widetilde \nabla_{b} \phi_{2j} - 4 \widetilde \nabla_{b} K^{cai}
K{_{c}}^{bj}\phi_{1i}\phi_{2j} - 4 K^{cai}K{_{c}}^{bj} \widetilde
\nabla_{b}\phi_{1i}\phi_{2j}] \nonumber \\ \!\!\!\ & + & \!\!\!\
\beta [-2 K^{cdi}K{_{cd}}^{j}\phi_{1i} \widetilde \nabla^{a}
\phi_{2j} + 2\widetilde \nabla^{a} K^{cdi}
K{_{cd}}^{j}\phi_{1i}\phi_{2j} + 2 K^{cdi} \widetilde \nabla^{a}
K{_{cd}}^{j} \phi_{1i}\phi_{2j} \nonumber \\ \!\!\! & + & \!\!\!
2K^{cdi} K{_{cd}}^{j} \widetilde \nabla^{a} \phi_{1i}\phi_{2j}].
\end{eqnarray}
In this manner, equation (37) implies that the
operator $P^{ij}$ is self-adjoint,  and considering that $\phi_{1i}$, and $\phi_{2j}$ correspond to
solutions of the equation (21) ($P^{ij} \phi_{1j}=P^{ij} \phi_{2j}=0$),  $j^{a}$ given in equation (38) is
a world sheet covariantly conserved
\begin{equation}
\nabla_{a}j^{a}=0.
\end{equation}
In the next section, we will compare the expression (38) with the
variation of the symplectic potential given in equation (19) on
the phase space, and we will demostrate that are exactly  the
same.
\newline
\newline
\noindent \textbf{V. The Witten phase space for [DNG-GB] strings and the Symplectic Structure on Z}\\[1ex]
In according to \cite{1}, in a given physical theory, the
classical phase space is the space of solutions of the classical
equations of motion, which corresponds to a manifestly covariant
definition, and on such phase space we can construct a covariant
and gauge invariant symplectic structure. The basic idea to
construct a symplectic structure on the space phase is to describe
Poisson brackets of the theory in terms of it, instead of choosing
$p's$ and
$q's$. \\
Based in the last paragraph, the Witten phase space for [DNG-GB]
p-branes, is the space of solutions of the equation (14)
\begin{eqnarray}
\nonumber  \sigma K^{i} +  2\beta G_{ab}K^{abi}=0,
\end{eqnarray}
but, for [DNG-GB] strings ($G_{ab}=0$) is the set of solutions of
the equation (18)
\begin{eqnarray}
\nonumber K^{i}=0,
\end{eqnarray}
and we shall call $Z$, and on this phase space we will construct a
symplectic structure. We can notice that the phase space for [DNG]
strings \cite{2,5} is the same for [DNG-GB] strings, equation
(18), but the corresponding symplectic structures, that are in the
transition of the regimens classical and quantum, will be
different, as we shall see below. In the literature, using a
conventional canonical formulation to quantize the [DNG-GB]
strings from the corresponding classical dynamics, equation (18),
the same results are obtained whether we include the topological
term or not, but in this scheme of quantization there is an
important contribution of such term as in path integral formalism,
where  such term has a relevant contribution weighting the
different topologies in the sum
over world surfaces.\\
Thus, following to \cite{2,5}, we can identify the scalar fields
$\phi^{i}$, $\phi^{a}$ as one-forms on $Z$ and therefore are
anticommutating objects:
$\phi^{i}\phi^{j}=-\phi^{j}\phi^{i}$, and  $\phi^{a}\phi^{b}=- \phi^{b}\phi^{a}$. Additional, in \cite{3,5} the vector field
$\delta=n^{i}\phi_{i}$ is identified as the exterior derivative on
$Z$, but it is incomplete, because of the exterior derivative on
$Z$ changes when we consider the importance of the tangential
deformations, and becomes to be
\begin{equation}
\delta =n^{i}\phi_{i} + e^{a}\phi_{a},
\end{equation}
since it is the correct exterior derivative which satisfies
\begin{equation}
\delta ^{2}=(n^{i}\phi_{i} +
e^{a}\phi_{a})(n^{j}\phi_{j} + e^{b}\phi_{b})=0,
\end{equation}
which vanishes because of the commutativity of the zero-forms
$n^{i}$, $e^{a}$ and the anticommutativity of the one-forms
$\phi^{i}$, $\phi^{a}$ on $Z$ \cite{2,5}.\\
In this manner, if we calculate the variation of symplectic
potential given in equation (19), we obtain
\begin{equation}
\delta \Psi^{a}=\textbf{D} \Psi^{a}=   \sqrt{-\gamma} j'^{a},
\end{equation}
with $j'^{a}$  given by
\begin{eqnarray}
 \nonumber j'^{a} \!\!\ & = & \!\!\
\sigma \phi_{i}\widetilde \nabla^{a}\phi^{i}- 4\beta  K^{bci}
\widetilde \nabla_{b} K{_{c}}^{aj} \phi_{i}\phi_{j}- 4\beta
K^{bci} K{_{c}}^{aj} \phi_{i} \widetilde \nabla_{b}\phi_{j} + 2
\beta K^{cdi} K{_{cd}}^{j} \phi_{i} \widetilde
\nabla^{a} \phi_{j} \nonumber \\
\!\!\ & +  & \!\!\  2\beta  K^{cdi}\widetilde
\nabla^{a}K{_{cd}}^{j}\phi_{i}\phi_{j},
\end{eqnarray}
 where we have used equations (8), (9), (41), and we have gauged away the
$\phi^{a}$ terms, because of is identified as a diffeomorphism on the world-sheet \cite{2}.\\
 Now, we compare the two-form obtained in equation (43) with the symplectic
 current found in last section considering the self-adjointness of the linearized dynamics.
 For this purpose, we can set $\phi_{1i} = \phi_{2i}= \phi_{i}$, and use the antisymmetry of
this scalar field
 in (38) \cite{2,5}, to obtain
\begin{eqnarray}
\nonumber j^{a} \!\!\ & = & \!\!\  \sigma \phi_{i}\widetilde
\nabla^{a}\phi^{i}- 4\beta  K^{bci} \widetilde \nabla_{b}
K{_{c}}^{aj} \phi_{i}\phi_{j}- 4\beta K^{bci} K{_{c}}^{aj}
\phi_{i} \widetilde \nabla_{b}\phi_{j} + 2\beta  K^{cdi}
K{_{cd}}^{j} \phi_{i} \widetilde
\nabla^{a} \phi_{j} \nonumber \\
\!\!\! & + & \!\!\! 2\beta  K^{cdi}\widetilde
\nabla^{a}K{_{cd}}^{j}\phi_{i}\phi_{j},
\end{eqnarray}
where we can notice that corresponds exactly to the two-form
obtained taking the variation of the symplectic potential,
equation (43).\\
Thus, we can notice that in string theory, the [DNG] action with
the [GB] topological term has indeed a physically relevant
contribution on the symplectic current found in equation (44), due
to the proportional terms to the parameter $\beta$. If the
parameter $\beta$ vanish, we obtain the symplectic current found
for [DNG] action \cite{2,5}.\\
On the other hand, if we take $\sigma=0$ and considering the case
of string theory in the equation (14), the dynamics vanishes, and
we would not find apparently any physical motivation for including
the [GB] topological term in any action describing strings,
however, in this paper the proportional terms to the parameter
$\beta$ in the equations (19), (20) and (44) do not vanish. In
this manner, whereas there is not dynamics of the system but we
have a non trivial deformations dynamics (see equation (20) with
$\sigma=0$), we can construct a non trivial  symplectic structure
that
lives in the transition of the regimes quantum and classical, that we will utilize in futures works.\\
It is important to notice that the treatment present in \cite{3} is incomplete, because of the exterior derivative
employed is not correct. Whereas this work that is consistent with the
results obtained by the adjoint operators method, equation (37),
and by the variation of the
symplectic potential, equation (43).\\
With the previous results, we can define a two-form on $Z$ in
terms of equation (44), that will be our symplectic structure
\begin{equation}
\omega\equiv \int_\Sigma \sqrt{-\gamma}j^{a}d\Sigma_{a}=
\int_\Sigma \delta \Psi^{a}d\Sigma_{a} ,
\end{equation}
were $\Sigma$ is a Cauchy surface for the configuration of the string.\\
We can see that $\omega$ is an exact two-form and in particular is
closed due to $\delta$ is nilpotent, this is
\begin{equation}
 \delta \omega= \int_\Sigma \delta(\delta
 \Psi^{a})d\Sigma_{a} = \int_\Sigma \textbf{D}(\textbf{D}
 \Psi^{a})d\Sigma_{a}=0.
\end{equation}
Now, we will prove that the symplectic structure, found in
equation (44), is a gauge invariant. For this purpose, we observe
the degenerate directions on the phase space associated with the
gauge transformations of the theory, such degenerate directions
will be associated with spacetime infinitesimal diffeomorphisms
\begin{equation}
X^{\mu}\rightarrow X^{\mu} + \delta X^{\mu}.
\end{equation}
Thus, we shall show that our symplectic structure $\omega$ is
invariant under the transformation given in equation (47). For
this, we use the fact that $\omega$ is given in terms of the
fields $\phi^{i}=n{^{i}}_{\mu} \delta X^{\mu}$ and under the
transformation (47), is invariant, this is
\begin{equation}
\phi^{i}=n{^{i}}_{\mu} \delta [X^{\mu}+
\delta X^{\mu}]=\phi^{i},
\end{equation}
where the second term vanishes because of $\delta$ is
nilpotent; in this manner, we have showed that $\omega$ is a gauge invariant.\\
It is important to notice that $\delta X^{\mu}$, in the
definition of $\phi^{i}$, physically represents the infinitesimal
spacetime deformations of the embedding, whereas
$\delta X^{\mu}$ in the transformation (47) is a spacetime
diffeomorphism \cite{5}.\\
Therefore, $\omega$ is a non degenerate two-form on $Z$ for
[DNG-GB] string theory.
\newline
\newline
\newline
\noindent \textbf{VI. Conclusions and prospects}\\[1ex]
 As we have seen, although in string theory  the [GB] topological term that
we have added to the [DNG] action does not contribute to the
dynamics of the system, it has a nontrivial contribution on the
Witten covariant phase space, by identifying a symplectic
potential as in \cite{2}, and using a covariant description of the
canonical formalism, which is completely unknown in the
literature. Taking this into account, we have constructed a
covariant and gauge invariant geometrical structure for [DNG-GB]
strings, from which we shall study, for example, the relevant
symmetries of the system and construct the corresponds Poisson
brackets. \\
It is important to mention that the  quantization aspects for
[DNG] action in string theory is well known, concretely the
solutions for the dynamics are known in an explicit way in
the literature, which is crucial in the study of such aspects,
but other cases are not considered, for example,
the system taken in this paper ([DNG-GB] branes), because of is difficult to solve the equations of motion. However, in the case tried here
the equations of motion for [DNG] and [DNG-GB] in
string theory are the same (equation (18)), and specifically their solutions. In this manner, we can take advantage
of this fact and use the solutions known to the equations of
motion for [DNG] string, and the results presented here to reveal
explicitly the contribution of the topological term on the
quantization aspects of the theory under study,
which has not been considered in the literature, however, this is a future work.\\
In addition, it is important to mention that the same treatment
presented in this paper is applicable for the First Chern number,
that also is a topological invariant for the world
sheet sweep by out a string embedded in a  4- dimensional background spacetime, but the
calculation we shall develop in future works when we will require
explicitly.
\newline
\newline
\noindent \textbf{Acknowledgements}\\[1ex]
This work was supported by CONACYT. The author wants to thank R.
Cartas-Fuentevilla for conversations and the friendship that he
has offered me.
\\[1em]

\end{document}